\documentclass[twocolumn,amssymb,amsmath,
nofootinbib,tightenlines,showpacs,floatfix,
superscriptaddress,aps,prb]{revtex4-1}
\usepackage{amsfonts}
\usepackage{txfonts} 
\usepackage{amsbsy}
\usepackage{bm}%
\usepackage[colorlinks=true,linkcolor=blue,filecolor=blue,menucolor=yellow,urlcolor=blue,citecolor=blue,anchorcolor=blue]{hyperref}
\usepackage{graphicx}
\usepackage{times} 
\usepackage{dcolumn}

\begin{document}

\title{Proximity effects and triplet correlations in 
Ferromagnet/Ferromagnet/Superconductor nanostructures} 

\author{Chien-Te Wu}
\email{wu@physics.umn.edu}
\author{Oriol T. Valls}
\email{otvalls@umn.edu}
\altaffiliation{Also at Minnesota Supercomputer Institute, University of Minnesota,
Minneapolis, Minnesota 55455}
\affiliation{School of Physics and Astronomy, University of Minnesota, 
Minneapolis, Minnesota 55455}

\author{Klaus Halterman }
\email{klaus.halterman@navy.mil}
\affiliation{Michelson Lab, Physics
Division, Naval Air Warfare Center, China Lake, California 93555}

\date{\today}

\begin{abstract} 
We report the results of a study of 
superconducting proximity effects in clean
Ferromagnet/Ferromagnet/Superconductor (${\rm F_1F_2S}$) 
heterostructures, where
the pairing state in S is a conventional singlet $s$-wave.
We numerically find the self-consistent solutions of the Bogoliubov-de Gennes (BdG) equations
and use these solutions to calculate  the relevant physical quantities.
By linearizing the BdG equations, we obtain  the superconducting
transition temperatures $T_c$ as a function of the angle $\alpha$ between the exchange fields 
in $\rm F_1$ and $\rm F_2$. 
We find that the results for $T_c(\alpha)$
in ${\rm F_1F_2S }$ systems are
clearly  different from those in 
${\rm F_1 S F_2}$ systems, where  
$T_c$ monotonically increases with $\alpha$ and is highest
for antiparallel magnetizations.
Here, $T_c(\alpha)$ is in general a non-monotonic function, 
and often has a 
minimum near $\alpha \approx 80^{\circ}$. 
For certain values of the exchange field and layer thicknesses, 
the system  exhibits reentrant  
superconductivity with $\alpha$: 
it transitions from superconducting to normal, 
and then returns to a superconducting state again 
with increasing $\alpha$.
%
This phenomenon is 
substantiated by a calculation of the  condensation  energy. 
We compute, in addition to
the ordinary singlet pair amplitude, the induced odd triplet pairing amplitudes.
The results indicate 
a connection 
between equal-spin triplet pairing and
the singlet pairing state that characterizes $T_c$.
We find also that the induced triplet amplitudes can 
be very long-ranged in both the S and F sides
and characterize their range. 
We discuss the average density of states 
for both the magnetic  and the S regions, 
and its relation
to the pairing amplitudes and $T_c$.
The local magnetization vector, which exhibits reverse
proximity effects, is also investigated. 
\end{abstract}


\maketitle

\section{Introduction}

Superconducting proximity effects in ferromagnet/superconductor heterostructures (F/S)
have received much attention in the past few decades  both for  their important applications 
in spintronics\cite{cite:zutic} and because of the underlying physics\cite{cite:buzdin}. 
Although ferromagnetism and $s$-wave superconductivity are largely
incompatible because of
the opposite nature of the 
spin structure of their order parameters, 
they can still coexist in nanoscale F/S systems via  
superconducting proximity effects\cite{cite:buzdin,cite:hv1}. 
The fundamental feature of proximity effects in F/S heterostructures is 
the damped oscillatory behavior of the superconducting order parameter in 
the F regions\cite{cite:demler}.
Qualitatively, the reason  is that a spin singlet Cooper pair acquires a finite momentum 
when it encounters the exchange field as it enters the 
ferromagnet. This affects the momenta of 
individual electrons that compose the Cooper pairs, and modifies 
both ordinary and Andreev\cite{andreev} reflection.
The interference between the transmitted and reflected Cooper pair
wave functions in the F regions leads to an 
oscillatory behavior of the 
dependence of the superconducting transition 
temperature, $T_c$,  on the thickness $d_F$ of the
ferromagnet in F/S bilayers\cite{cite:buzdin,cite:radovic,cite:jiang}. 
Because of these oscillations the superconductivity may even disappear in a certain 
range of F thicknesses. 
Indeed, this reentrant superconductivity with 
geometry was theoretically predicted and experimentally 
confirmed.\cite{cite:khusainov,cite:garifullin,zdravkov,fominov,buzdin3,hv3,hv4} 

Another remarkable fact related to 
F/S proximity effects is that  triplet pairing correlations 
may be induced in F/S systems where S is
in  the ordinary $s$-wave pairing 
state.\cite{berg86,berg85,lof,cite:hv2,hv2p} 
These correlations can be long ranged, extending deep into both the F and S regions. 
The Pauli principle requires the corresponding condensate 
wavefunction (in the $s$-channel) 
to be 
odd in frequency\cite{berez} or time\cite{cite:hv2}. 
The magnetic inhomogeneity arising from the presence of the ferromagnet 
in F/S systems is responsible for this type of triplet pairing. 
The components of the triplet pairing correlations are restricted, because
of conservation laws, by the magnetic 
structure in the F layers: only the total spin projection corresponding to the 
$m=0$ component can be induced when the exchange fields
arising from the ferromagnetic structure are all aligned in the same direction,
while 
all three components ($m=0,\pm1$) can arise when the exchange fields
are not aligned. 
Because of the exchange fields, 
singlet pairing 
correlations decay 
in F with a short range decay length. 
On the other hand, the induced triplet pairing correlations can be long ranged, with their 
length scale being comparable to that of the usual slow decay associated with 
nonmagnetic metal proximity effects. 
Early experiments revealed a long range decay length in the differential resistance in
a ferromagnetic metallic wire (Co) which can be well explained within a framework 
that accounts for 
triplet pairing correlations.\cite{cite:giroud}
More recently, experimental observations  
of long range spin triplet supercurrents have been reported
in several multilayer systems, \cite{khaire,gu,sprungmann} 
and also in  
Nb/Ho bilayers.\cite{cite:robinson}
In the last case, the requisite magnetic inhomogeneity arises from 
the spiral magnetic structure inherent to the rare earth compound, Ho, 
which gives rise
also to oscillations\cite{usprl} in $T_c$. 
Other theoretical work\cite{cite:eschrig,eschrig2} 
in the semiclassical limit  shows that in the half metallic ferromagnet case
spin flip scattering at
the interface provides  a mechanism for conversion between a short range
singlet state and an ordinary (even in frequency or time)
triplet one in the
$p$-wave channel. This holds also\cite{grein} for strongly 
polarized magnets.

Both the short and long spatial range 
of the oscillatory singlet and odd 
triplet correlations in the
ferromagnetic regions permit
control over the critical temperature, $T_c$, that is, the switching on or off of
superconductivity. The long range propagation of
equal spin triplet correlations
in the ferromagnetic regions 
was shown to
contribute to a spin valve effect 
that varies with the relative magnetization in the F layers\cite{golubov}.
With continual interest in nonvolatile memories, a number of spin valve type
of structures have been proposed. These use 
various arrangements of S and F layers to turn superconductivity on or off.
Recent theoretical work suggests that when two 
ferromagnet layers are placed in direct contact and adjacent to a superconductor,
new types of
spin valves\cite{oh,golubov,karmin} or Josephson 
junctions\cite{cheng,berg86,knezevic} 
with interesting and unexpected  behavior can ensue.
For an  $\rm F_1 F_2 S$ superconducting memory device\cite{oh},  the
oscillatory decay of the singlet correlations can be manipulated by switching
the relative magnetization in the F layers from parallel to antiparallel by
application of an external magnetic field.
It has also been shown\cite{golubov} using
quasiclassical methods 
that for these $\rm F_1 F_2 S$ structures 
the critical temperature
can have a minimum at a relative magnetization angle
that lies between the parallel and antiparallel configuration. This is in
contrast with $\rm F_1 S F_2$ trilayers, where (as indicated by 
both\cite{cite:zhu,cite:tagirov,cite:buzdin4,buzdin3} 
theory and experiment\cite{cite:gu,cite:potenza,cite:moraru} )
the behavior of $T_c$ with relative angle is strictly monotonic, with
a minimum when the magnetizations are parallel and a maximum when antiparallel. 
For $\rm S F_1 F_2 S$ type structures, the exchange
field in the magnets can increase the Josephson current,\cite{berg86}
or, in the case on non-collinear alignment,\cite{knezevic} induce triplet correlations
and discernible signatures in the corresponding density of states.

Following up on this work, an 
${\rm F_1 F_2 S}$ spin switch was experimentally demonstrated\cite{leksin}
using ${\rm CoO_{x}/Fe1/Cu/Fe2/In} $ multilayers. 
Supercurrent flow through the sample was completely inhibited 
by changing the mutual orientation of the magnetizations
in the two adjacent F layers.
A related phenomenon was reported\cite{leksin2} for a similar multilayer spin valve,
demonstrating that the
critical temperature can be higher for parallel orientation
of relative magnetizations.
A spin valve like effect was also experimentally realized\cite{west,nowak} in $\rm Fe V$
superlattices, where antiferromagnetic coupling between the Fe layers permits gradual 
rotation of the relative magnetization direction in the $\rm F_1$ and
$\rm F_2$ layers. 

As already mentioned, the $T_c(\alpha)$  behavior  in the $\rm F_1F_2S$  geometry 
is in stark 
contrast to that 
observed in the 
more commonly studied
spin switch structures involving
$\rm F_1SF_2$  configurations. There, as the angle $\alpha$ between the
(coplanar) magnetizations
increases from zero (parallel, P, configuration) to $180^\circ$ (antiparallel, AP, 
configuration) $T_c$ increases monotonically.
For these systems it has been demonstrated too 
that under many conditions they can be made to
switch from a superconducting state (at large $\alpha$) to a normal one \cite{hv4,cite:gu} 
in the P configuration, by flipping the  magnetization
orientation in one of the F layers. The
AP state however is robust: it is always the lowest energy state regardless of
relative strength of the ferromagnets, interface scattering, and geometrical variations.
The principal reason for this stems from the idea that the average exchange field overall
is smaller for the AP relative orientation of
the magnetization. 
Early experimental data on $T_c^{AP}$ and $T_c^P$, where $T_c^{AP}$ and $T_c^P$ 
are the transition temperatures for 
the AP and P configurations, was obtained in CuNi/Nb/CuNi\cite{cite:gu}.
There $\Delta T_c \equiv T_c^{AP}-T_c^P>0$,  
was found to be about 6 mK.
Later, it was found that $\Delta T_c$ can be as large as 41 mK in Ni/Nb/Ni trilayers\cite{cite:moraru}.
Recently, the angular dependence of $T_c$ of $\rm F_1SF_2$ systems was also measured 
in CuNi/Nb/CuNi trilayers and its monotonic behavior found
to be in good agreement with theory\cite{cite:zhu}.
In addition to the experimental work, the thermodynamic properties of $\rm F_1SF_2$ 
nanostructures were studied quasiclassically by solving the Usadel 
equations\cite{cite:buzdin4}. It was seen that these properties
are strongly dependent on the mutual orientation of the F layers. The difference in the free
energies of the P and AP states can be of the same order of magnitude as 
the superconducting condensation energy itself.
In light of the differences between $\rm F_1F_2S$ and $\rm F_1SF_2$,
it appears likely that
a full microscopic theory is needed that
accounts for the geometric interference effects and
quantum interference effects that are present due to the various scattering processes.

\begin{figure}
\includegraphics[width=0.5\textwidth] {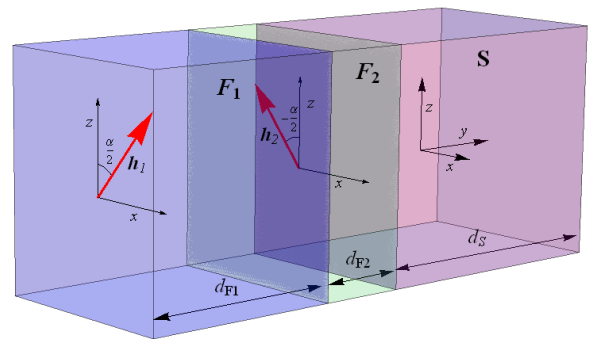} 
\caption{(Color online) Schematic of the F$_1$F$_2$S trilayer. 
The outer 
ferromagnetic layer 
$\rm F_1$ has a  magnetization oriented at an angle $\alpha/2$ in the $x-z$ plane, while
the inner 
ferromagnet, $\rm F_2$, has a magnetization orientation at an
angle $-\alpha/2$  in the $x-z$ plane. All relevant widths are labeled.
}
\label{fig0}
\end{figure}

In this paper, 
we consider several 
aspects to the proximity effects
that arise in $\rm F_1 F_2 S$ spin switch nanostructures:
We consider  arbitrary
relative orientation of the magnetic moments in the two F
layers and study both the singlet and the induced odd 
triplet correlations in the clean limit through a fully
self-consistent solution of the microscopic 
Bogoliubov-de Gennes (BdG) equations. 
We also calculate the critical temperature by solving
the linearized BdG equations.
As a function of the angle $\alpha$,  
it is often non-monotonic, possessing a minimum that lies approximately midway between
the parallel and antiparallel configurations. Reentrant
behavior occurs  when this minimum drops
to zero. 
We find that there are induced odd 
triplet correlations and we
study their behavior.  
These correlations are found to be often
long ranged in both the S and F regions.
These findings are consistent with the single particle behavior
exhibited by the density of states and magnetic moment in these structures.

\section {Methods} \label{methods}

We consider a trilayer ${\rm F_1 F_2 S}$ structure  
infinite in the $x-z$ plane, and with total length $d$ in the $y$ direction, 
which is normal
to the interfaces. The inner ferromagnet layer ($\rm F_2$) of width $d_{F2}$
is adjacent to the outer 
ferromagnet ($\rm F_1$)
of
width $d_{F1}$, and the superconductor has width $d_S$ (see Fig.~\ref{fig0}). 
The magnetizations in the  ${\rm F_1}$ and ${\rm F_2}$ layers
form angles $\alpha/2$ and $-\alpha /2$, respectively, with the axis of quantization 
$z$. 
The superconductor is of the conventional $s$-wave type. 
We describe the magnetism of
the F layers by an effective exchange field ${\bf h}(y)$ that 
vanishes in the S layer. 
We assume that interface scattering barriers are negligible,
in particular that there is no interfacial spin flip scattering. 
Our methods are described 
in Ref.~\onlinecite{cite:hv2,hv2p}  and  details that are not pertinent to 
the specific  problem we consider here will not be repeated. 

To accurately describe the behavior of the quasiparticle ($u_{n\sigma}$) and
quasihole ($v_{n\sigma}$) amplitudes with spin $\sigma$, 
we use the Bogoliubov-de Gennes\cite{bdg} (BdG) formalism. 
In our geometry, the BdG equations 
can be written down after  a few steps\cite{hv2p}  
in the quasi-one-dimensional form: 

\begin{align}
&\begin{pmatrix} 
{\cal H}_0 -h_z&-h_x&0&\Delta(y) \\
-h_x&{\cal H}_0 +h_z&\Delta(y)&0 \\
0&\Delta(y)&-({\cal H}_0 -h_z)&-h_x \\
\Delta(y)&0&-h_x&-({\cal H}_0+h_z) \\
\end{pmatrix}
\begin{pmatrix} 
u_{n\uparrow}(y)\\u_{n\downarrow}(y)\\v_{n\uparrow}(y)\\v_{n\downarrow}(y)
\end{pmatrix} \nonumber \\
&=\epsilon_n
\begin{pmatrix}
u_{n\uparrow}(y)\\u_{n\downarrow}(y)\\v_{n\uparrow}(y)\\v_{n\downarrow}(y)
\end{pmatrix}\label{bogo2},
\end{align}
where ${\cal H}_0$ is the usual single particle Hamiltonian,
${\bf h}(y)=(h_x(y),0,h_z(y))$ is the exchange field in the F layers, 
$\Delta(y)$ is the pair potential, taken to be real,  and  
the  wavefunctions  $u_{n \sigma}$ and $v_{n \sigma}$ are the standard
coefficients that appear when the usual field operators 
$\psi_\sigma$ are expressed in terms of a Bogoliubov transformation:
\begin{equation}
\label{bv}
\psi_{\sigma}({\bf r},t)=\sum_n \left(u_{n\sigma}({\bf r})\gamma_n e^{-i \epsilon_n t} +\eta_\delta v_{n\sigma}({\bf r})
\gamma_n^\dagger e^{i \epsilon_n t}\right),  
\end{equation}
where $\eta_\delta\equiv 1(-1)$ for spin down (up). 
We must include all four spin components since the
exchange field in the ferromagnets destroys the spin rotation invariance.

To ensure that the system is in an, at least 
locally, thermodynamically  stable state, 
Eq.~(\ref{bogo2}) must be solved jointly with the
self consistency condition for the pair potential:
\begin{equation}  
\label{del} 
\Delta(y) = \frac{g(y)}{2}{\sum_n}^\prime    
\bigl[u_n^\uparrow(y)v^\downarrow_n (y)+
u_n^\downarrow(y)v^\uparrow_n (y)\bigr]\tanh\Bigl(\frac{\epsilon_n}{2T}\Bigr), \,
\end{equation} 
where the primed sum is over eigenstates corresponding to 
positive energies smaller than or equal to  the ``Debye'' 
characteristic energy cutoff $\omega_D$, and  
$g(y)$ is the superconducting coupling parameter that is a constant 
$g_0$ in the intrinsically superconducting
regions and zero elsewhere. 

With the above assumptions on interfacial scattering, 
the triplet correlations are odd in time, in agreement with the Pauli
principle and hence vanish at $t=0$. Therefore 
we will consider the time dependence of 
the triplet correlation functions,
defined\cite{cite:hv2} in terms of the usual field operators as, 
\begin{subequations}
\label{alltriplet}
\begin{align}
f_0 ({\bf r},t) &\equiv \frac{1}{2}[\langle \psi_\uparrow({\bf r},t)\psi_\downarrow({\bf r},0) \rangle+
\langle \psi_\downarrow({\bf r},t)\psi_\uparrow({\bf r},0) \rangle],
\label{f0def}
\\
f_1 ({\bf r},t) &\equiv \frac{1}{2}[\langle \psi_\uparrow({\bf r},t)\psi_\uparrow({\bf r},0) \rangle-
\langle \psi_\downarrow({\bf r},t)\psi_\downarrow({\bf r},0) \rangle],
\label{f1def}
\end{align}
\end{subequations}

These 
expressions can be conveniently written
in terms of the quasiparticle amplitudes:\cite{cite:hv2,hv2p} 
\begin{subequations}
\label{alltripleta}
\begin{align}
f_0 (y,t) & = \frac{1}{2} \sum_n \left[ u_{n\uparrow} (y) v_{n\downarrow}(y)-
u_{n\downarrow}(y) v_{n\uparrow} (y) \right] \zeta_n(t), 
\label{f0defa} \\
f_1 (y,t) & = \frac{1}{2} \sum_n \left[ u_{n\uparrow} (y) v_{n\uparrow}(y)+
u_{n\downarrow}(y) v_{n\downarrow} (y) \right] \zeta_n(t),
\label{f1defa}
\end{align}
\end{subequations}
where $\zeta_n(t) \equiv \cos(\epsilon_n t)-i \sin(\epsilon_n t) \tanh(\epsilon_n /(2T))$,
and  {\it all} positive energy states are in general summed over.

Besides the pair potential
and the triplet amplitudes, we can also determine various physically relevant single-particle quantities.
One such important quantity is the local magnetization, which can 
reveal details of 
the well-known (see among many
others, Refs. \onlinecite{fryd,koshina,usold, bergeretve}) reverse 
proximity effect: the penetration of the magnetization into S. 
The local magnetic moment ${\bf m}$ will depend on
the coordinate $y$  and it will
have in general both $x$ and $z$ components, ${\bf m} = (m_x,0,m_z)$.
We define 
${\bf m}=-\mu_B\langle\sum_{\sigma} \Psi^\dagger {\bm \sigma} \Psi \rangle$, 
where $\Psi^\dagger \equiv (\psi_\uparrow, \psi_\downarrow)$. 
In terms of the quasiparticle amplitudes calculated 
from the self-consistent BdG equations 
we have,
\begin{subequations}
\label{mm}
\begin{align}
m_x(y) =& -2 \mu_B \sum_{n}\biggl\lbrace u^{\uparrow }_{n}(y) u^\downarrow_{n}(y) 
f_n 
-v^\uparrow_{n}(y) v^{\downarrow }_{n}(y)
(1- f_n)\biggr\rbrace, \\ 
m_z(y) =& -\mu_B \sum_{n}\biggl\lbrace (|u^\uparrow_{n}(y)|^2- |u^\downarrow_{n}(y)|^2) f_n \nonumber \\
&+(|v^\uparrow_{n}(y)|^2- |v^\downarrow_{n}(y)|^2) (1- f_n)\biggr\rbrace, 
\end{align}\end{subequations}
where $f_n$ is the Fermi function of $\epsilon_n$ and $\mu_B$ is the Bohr magneton. 

A very useful tool in the study of these systems is 
tunneling spectroscopy, 
where information, measured by an STM, can reveal the local DOS (LDOS). Therefore
we have computed here also the LDOS $N(y,\epsilon)$ as a function of $y$. We have  
$N(y,\epsilon) \equiv N_\uparrow (y,\epsilon) + N_\downarrow (y,\epsilon)$, where,
\begin{align}
\label{dos}
N_\sigma(y,\epsilon) = \sum_n [u_{n\sigma}^2(y) \delta(\epsilon-\epsilon_n)+ v_{n\sigma}^2(y) \delta(\epsilon+\epsilon_n)], 
\quad \sigma = \uparrow, \downarrow. 
\end{align}

The transition temperature  can be calculated for our system by finding the 
temperature at which the pair potential vanishes. It is much more efficient, however, to 
find $T_c$ by linearizing\cite{bvh}
the self-consistency equation   near the
transition, leading
to the form 
\begin{align}
\Delta_i=\sum_q J_{iq}\Delta_q, 
\end{align}
where the $\Delta_i$ are expansion coefficients 
of the position dependent pair potential  in
the chosen  basis and the  $J_{iq}$ are the appropriate 
matrix elements with respect to the same basis. The somewhat lengthy 
details of their evaluation 
are given in Ref.~\onlinecite{bvh}. 

To evaluate the free energy, $F$, of the self-consistent
states we use the convenient expression,\cite{kos}
\begin{align}
F = -2T \sum_n \ln \left[2 \cosh \left(\frac{\epsilon_n}{2T}\right )
\right]+\left \langle \frac{\Delta^2(y)}{g(y)} \right \rangle_s,
\label {fe}
\end{align}
where here $\langle \ldots \rangle_s$ denotes spatial average. 
The condensation free energy, $\Delta F$, is defined as $\Delta F \equiv F_S - F_N$,
where $F_S$ is the free energy of the superconducting
state and $F_N$ is that of the non-superconducting system. 
We compute $F_N$ by setting $\Delta \equiv 0$ in Eqs.~(\ref{bogo2}) and (\ref{fe}). 

\begin{figure}
\includegraphics[width=0.5\textwidth] {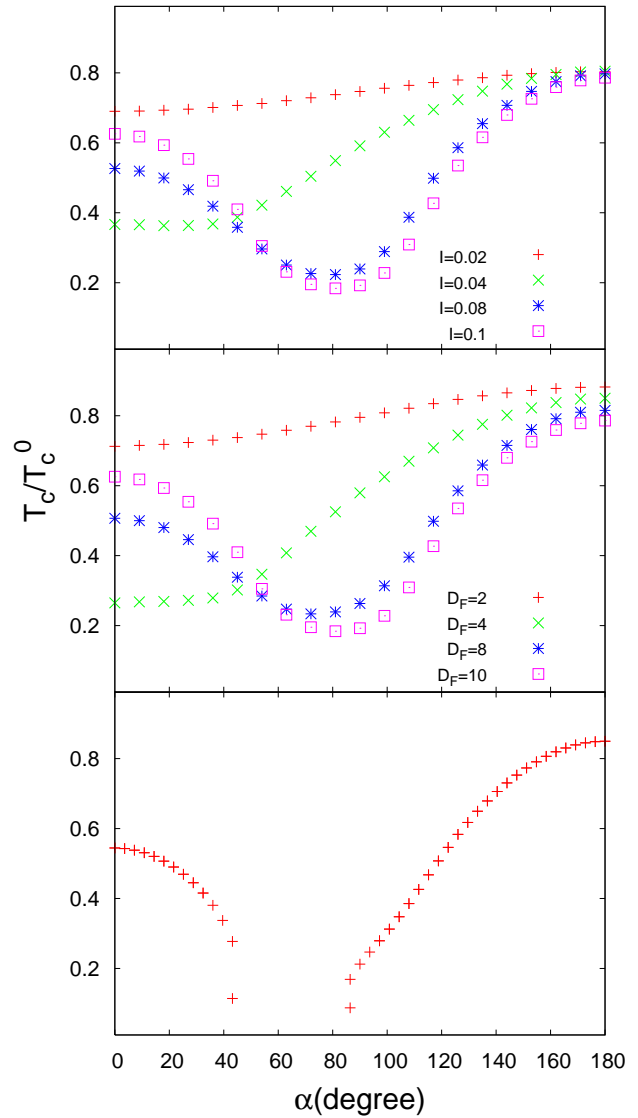} 
\caption{(Color online) Calculated transition temperatures $T_c$, normalized to 
$T_c^0$. In this figure the two F layers are
identical, $D_{F1}=D_{F2}\equiv D_F$ and $I_1=I_2\equiv I$. In the top panel
this ratio is shown
vs $\alpha$ for different exchange fields at
$D_{F}=10$ . In the middle panel 
the same ratio is plotted
again  vs $\alpha$ for different values of $D_F$ at
$I=0.1$. In the bottom panel  $T_c$
vs $\alpha$ is shown for  $D_{F}=6$ and $I=0.15$, a case where
reentrance with angle occurs.}
\label{fig1}
\end{figure}

\section {Results}

In presenting our results below we measure all lengths 
in units of the inverse of $k_F$ and denote by a capital letter
the lengths thus measured. Thus for example $Y\equiv k_F y$. 
The exchange field strength is measured by the dimensionless
parameter $I\equiv h/E_F$ where $E_F$ is the band width 
in S and $h$ the magnitude of the exchange 
field ${\bf h}$. In describing the two F layers
the subscripts 1 and 2 denote (as in Fig.~\ref{fig0}) the outer and inner layers respectively.
Whenever the two F layers are identical in some respect the
corresponding quantities  are given without an index: thus $I_2$
would refer to the inner layer while simply $I$ refers to both when this
is appropriate. 
We study a relatively wide range of thicknesses $D_{F1}$ for the outer
layer but there would be 
little purpose in studying thick inner layers
beyond the range of the standard singlet proximity effect in the magnets. 
In all cases we have assumed a superconducting
correlation length corresponding to $\Xi_0=100$ 
and measure all temperatures in units of $T_c^0$, the transition
temperature of {\it bulk} S material. The quantities $\Xi_0$ and 
$T_c^0$ suffice to characterize the BCS singlet material we consider.
We use $D_S=1.5 \Xi_0$ unless, as otherwise indicated, a larger 
value is needed to study penetration
effects. Except for the
transition temperature itself, 
results shown were obtained in the low 
temperature limit. For the triplet amplitudes, dimensionless
times $\tau$ are defined as $\tau \equiv \omega_D t$. Except for this
definition, the cutoff frequency plays no 
significant role in the results.

\begin{figure}
\includegraphics[width=0.5\textwidth] {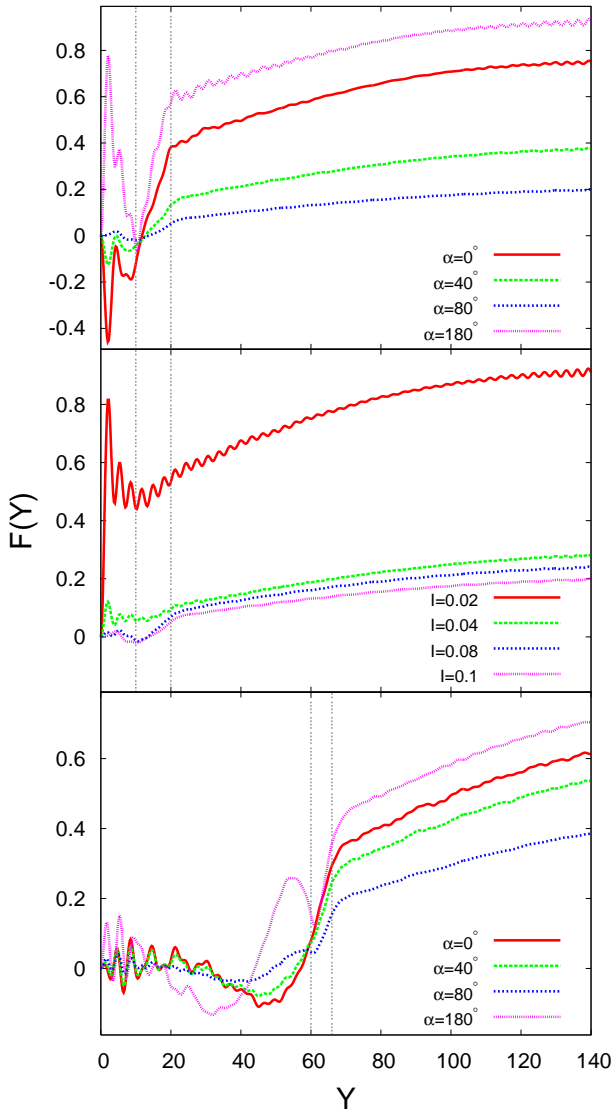}
\caption{(Color online)  Calculated singlet pair amplitude $F(Y)$, normalized to 
its  value in bulk
S material, plotted vs $Y\equiv k_Fy$. 
In the top panel, results are shown for different $\alpha$ at
$I=0.1$ and $D_F=10$.
The 
central panel depicts results
for the same $D_F$, and illustrates the effect of different 
magnetic strengths, $I$, at  fixed  $\alpha=80^\circ$. 
The bottom panel shows $F(Y)$ for different $\alpha$ as in the top panel,
except for a structure of differing magnet thicknesses:
$D_{F1} = 60$, and $D_{F2} = 6$.  The dashed vertical lines represent in each case the 
location of the ${\rm F_1F_2}$  and  ${\rm F_2S}$ interfaces.} 
\label{fig2}
\end{figure}

\begin{figure*}
\includegraphics[width=1\textwidth] {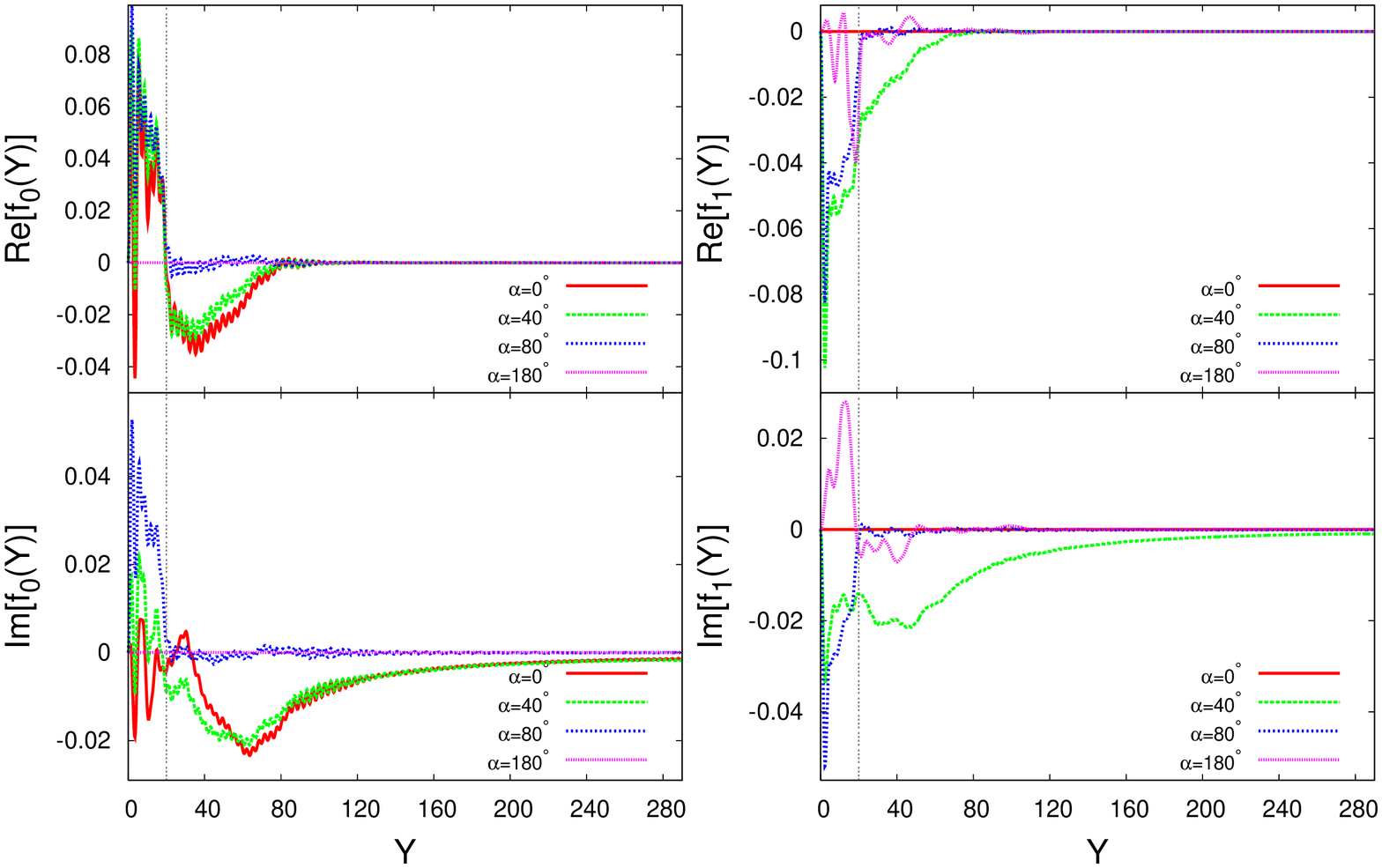}
\caption{(Color online) 
The Real and Imaginary parts
of the normalized triplet amplitudes $f_0$ and $f_1$
(see text) plotted vs $Y$ for a sample with
$D_{F}=10$, $D_S=300$ and $I=0.1$, at dimensionless time $\tau=4.0$. Results
are plotted for different values of $\alpha$ as indicated. See text
for discussion. Vertical lines indicate, in this and
the next three figures, the ${\rm F_2S}$ interface. For clarity, the 
$\rm F_1F_2$ interface is not included. 
}
\label{fig3}
\end{figure*}

\subsection{Transition Temperature}

The transition temperature $T_c$ is calculated directly from the 
linearization method described in Sec.~\ref{methods}. Some of the results
are shown in Fig.~\ref{fig1}. In this figure we have taken both F layers
to be identical and hence both relatively thin. All three
panels in the figure display  $T_c$, normalized to 
$T_c^0$, 
as a function
of the angle $\alpha$. The figure dramatically displays, as anticipated
in the Introduction, that as opposed to ${\rm F_1SF_2}$ trilayers, $T_c$ does not
usually, in our present case, monotonically increase 
as $\alpha$ increases from 0 to $180^\circ$, but on the contrary it has 
often a minimum at a value of $\alpha$ typically below $90^\circ$.

The top panel, which  shows results for several intermediate 
values of $I$ 
with  $D_F=10$, illustrates the above statements.  $T_c$ is found in this case
to be monotonic only at the smallest value of $I$ ($I=0.02$) considered. The
non-monotonic behavior starts to set in at around $I=0.04$ and then it continues,
with the minimum $T_c$ remaining at about $\alpha=80^\circ$. This is not
a universal value: we have found that for other geometric and material parameters 
the position of the minimum can be lower or higher. In the middle panel 
we consider a fixed value of 
$I=0.1$ and several values of $D_F$. This
panel makes another important point: the four curves plotted in the top
panel and the four ones in this panel correspond to identical
values of the product $D_F I$. 
The results, while not exactly the
same, are extremely similar and confirm that the oscillations in $T_c$
are determined by the overall periodicity of the Cooper pair amplitudes
in $F$ materials as determined by the difference between up and down
Fermi wavevectors, which is approximately
proportional\cite{hvlast} to $1/I$ in the range of $I$ shown. 

In the lowest 
panel of the figure we show that reentrance with $\alpha$
can occur in these structures. The results there are for $D_F=6$ and 
at $I=0.15$, a value a little larger than that considered in the other 
panels. While
such reentrance is not the rule, we have found that it is not an
exceptional situation either: the minimum in $T_c$ 
at intermediate $\alpha$ can simply drop 
to zero, resulting in reentrance. 
The origin of this reentrance stems
from the presence of triplet correlations due to the inhomogeneous magnetization
and the 
usual $D_F$ reentrance in F/S 
bilayers\cite{cite:buzdin,cite:khusainov,fominov,buzdin3,hv3,hv4,bvh}, 
that is, the periodicity of the pair amplitudes mentioned above.

\begin{figure*} 
\includegraphics[width=1\textwidth] {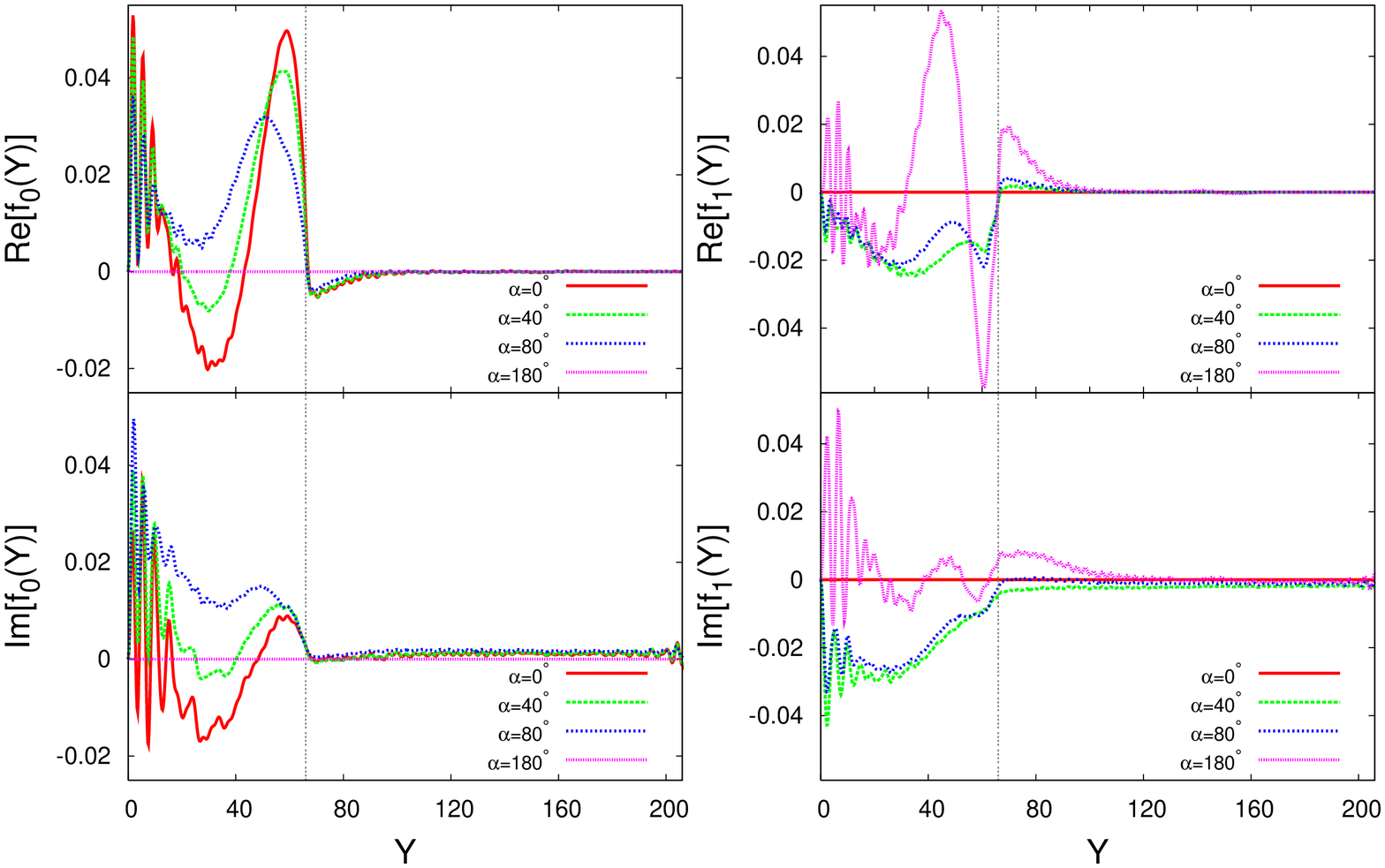}
\caption{(Color online) The real and imaginary parts of
the triplet amplitudes, plotted as in the previous figure
except that the sample has
$D_{F1}=60, D_{F2}=6$ and $D_S=150$. See text for discussion.
}
\label{fig4}
\end{figure*}

\subsection{Pair amplitude: singlet}

We turn now to the behavior of the standard, singlet pair amplitude $F(y)$,
defined as usual via $\Delta(y) \equiv g(y) F(y)$
and Eq.~(\ref{del}), 
as evaluated from the self consistent calculations described
in Sec.~\ref{methods}.
The behavior of $F(y)$  is rather straightforwardly described and has some 
features representative of conventional proximity effects found
in other ferromagnet-superconductor configurations,
such as F/S or ${\rm F_1SF_2}$ structures. 
An example is shown
in Fig.~\ref{fig2}, where that spatial behavior of $F(y)$ is shown 
for a few cases of  exchange fields differing in orientation and magnitude, 
as well as ferromagnet widths. 

The top panel shows results 
for $F(Y)$ as a function of position, at $I=0.1$ and for several
values of $\alpha$, at $D_F=10$.  
We see that in the S layer, the pair amplitude rises steadily
over a length scale of order of the correlation length. The variation
of the overall amplitude in S with $\alpha$ reflects that of the transition
temperature, as was depicted for this case by the (purple) squares in the top panel of 
Fig.~\ref{fig1}. One sees that the   non-monotonic trends observed 
in the critical temperature correlate well 
with the zero temperature pair amplitude
behavior. 
In the F layers, we 
observe 
a more complicated behavior and oscillations with an overall smaller amplitude. 
These oscillations are characteristic of 
conventional F/S proximity effects, which in this case appear
somewhat chaotic because of reflections and interference at the ${\rm F_1F_2}$
and end 
boundaries. This irregular spatial behavior is
also due to the chosen value of $I$ and the characteristic spatial 
periodicity  $\approx 2 \pi /I$ not matching $D_F$.  
These geometric effects can in some cases,
result in  the 
amplitudes of the singlet pair oscillations in ${\rm F_2}$ 
exceeding those in the superconductor near the interface.
 
In the central panel
results for several values of $I$ and the same
geometry as the top one 
are shown where the typical location of the minimum in $T_c$ may occur 
at a relative magnetization angle of $\alpha\approx 80^\circ$. 
We see that for the case $I=0.02$, where $T_c$ is high and monotonic with $\alpha$, 
singlet correlations are significant and they are 
spread throughout the entire structure.
This is consistent with the top panel of Fig.~\ref{fig1},
where the critical temperature is highest, and increases only slightly 
with $\alpha$. 
For the 
other values of
$I$, there is a strong $T_c$ minimum
near $\alpha=80^\circ$ and consequently, the pair
amplitude is much smaller. 
The weakening of the superconductivity in S
inevitably leads to its weakening in the F layers. 

The bottom panel demonstrates 
how the pair amplitude in the structure becomes modified when
$\alpha$ is varied, in a way similar to the top panel, 
except in this case the inner layer is thinner with $D_{F2} = 6$
and the outer layer is thicker with $D_{F1} = 60$. Comparing the top
and bottom panels, we see that clearly geometric effects
can be quite influential on the
spatial behavior of singlet pairing correlations. 
In this case the ${\rm F_2}$ layer is too thin for $F(Y)$
to exhibit oscillations within it. 

\begin{figure*} 
\includegraphics[width=1\textwidth] {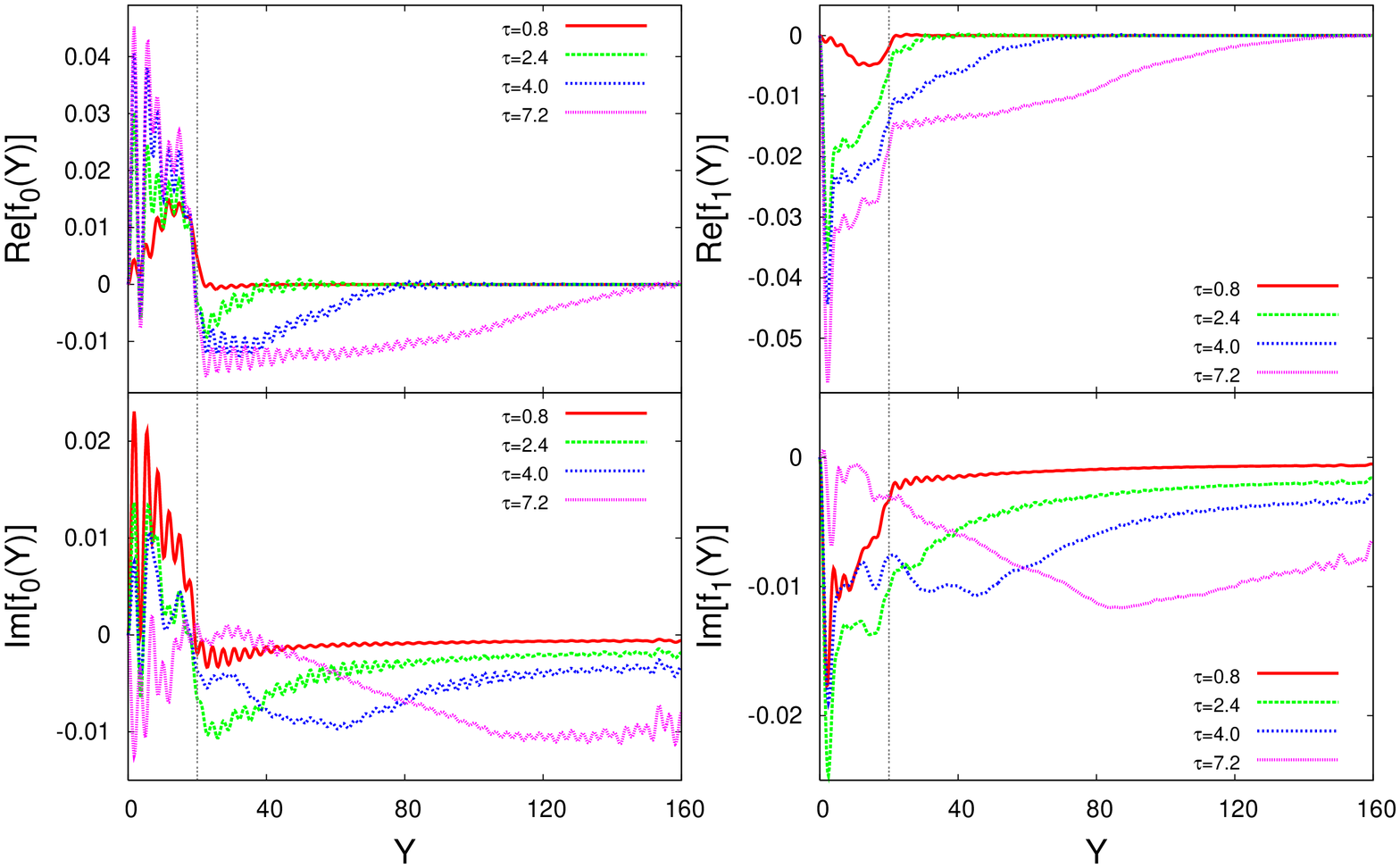}
\caption{(Color online) The real and imaginary parts
of the triplet amplitudes plotted vs $Y$ for the
same parameter values and conventions as in Fig.~\ref{fig3}, at fixed 
$\alpha=40^\circ$ for several values of $\tau$ as indicated.
\label{fig5}
}
\end{figure*}

\subsection{Triplet amplitudes}

In this subsection, 
we discuss the induced triplet pairing correlations in our systems.
As mentioned in the Introduction, the triplet pairing correlations may coexist with
the usual singlet pairs in F/S heterostructures and their behavior 
is in many ways quite different: in particular
the characteristic proximity
length can be quite large. As a function of the
angle $\alpha$ the possible existence of the different triplet
amplitudes is restricted\cite{cite:hv2,hv2p} by conservation laws.
For instance, at $\alpha=0$ (parallel exchange fields) the 
$m=\pm 1$ component along our $z$ axis of quantization, $f_1(y,t)$,
must identically vanish, while $f_0$ is allowed. This is because
at $\alpha=0$  the $S_z$  component of the total Cooper pair spin
is conserved, although the total spin quantum number $S$ is not. Neither
quantity is conserved for arbitrary $\alpha$. For  directions
other than $\alpha=0$ 
restrictions arising from the symmetry properties 
can be inferred\cite{hv2p} 
most easily 
by projecting onto the $z$ axis the quasiparticle amplitudes  
along a different axis in the $x-z$ plane via 
a unitary spin rotation operator, $U$:
\begin{align}
{U}(\varphi)=
\cos(\varphi/2) \hat{\bf 1}\otimes\hat{{\bf 1}}-i\sin(\varphi/2)\rho_z\otimes\sigma_y,
\end{align}
where $\varphi$ is measured from the $z$-axis, and
$\rho_z$ is a Pauli-like operator, acting
in particle-hole space.
For the anti-parallel case, $\alpha=180^\circ$, we have, 
following from the operation of spin rotation above,
the inverse property that
only $f_1$ components can be induced. In addition, the Pauli
principle requires all triplet amplitudes to vanish at $t=0$.
We note also that with the usual phase convention taken here,
namely that the singlet amplitude is real, the triplet amplitudes may
have, and in general they do have, both real and imaginary parts.
The results of triplet amplitudes shown here are calculated at zero 
temperature and are normalized to the singlet pair amplitude of a bulk S material. 

\begin{figure*}
\includegraphics[width=1\textwidth] {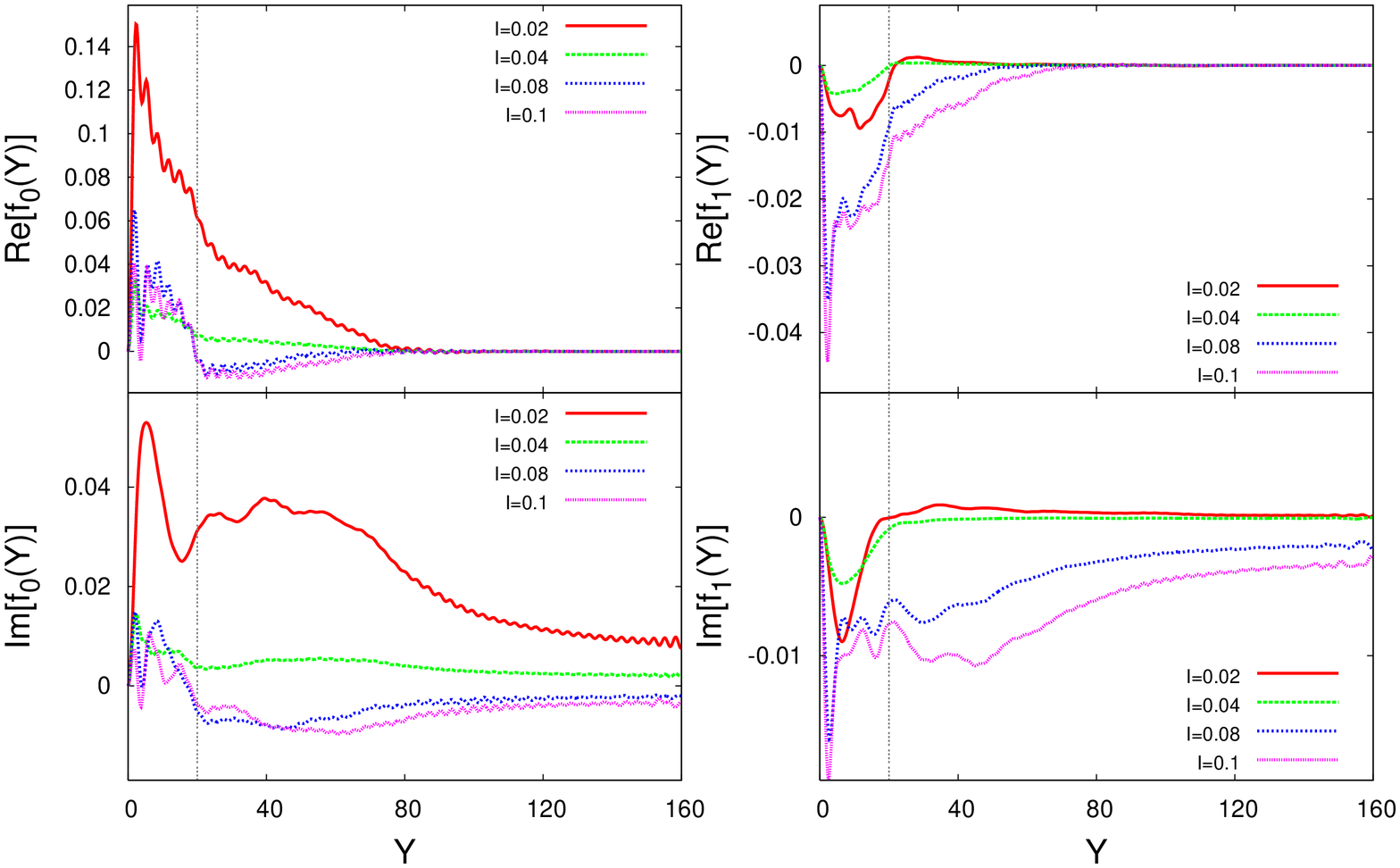}
\caption{(Color online) The triplet amplitudes $f_0$ and $f_1$
plotted as a function of position at fixed $\alpha=40^\circ$  and
$\tau=4$ for several values of $I$. We have here $D_F=10$, and $D_S=150$. 
}
\label{fig6}
\end{figure*}

First, we present in Fig.~\ref{fig3} the case of a  
thick S layer ($D_S=300$)  with two thin F layers ($D_F=10$). 
The two F layers have exchange fields of identical magnitude, corresponding to $I=0.1$,
and the angle $\alpha$ is varied. The dimensionless time
chosen is $\tau=4$, the behavior is characteristic of all times
in the relevant range. Of course the results at $\tau=0$ are
found to vanish identically.
As observed in this figure, the results for $f_1$ vanish
at $\alpha=0$ and those for $f_0$ at $\alpha=\pi$ in agreement
with the conservation law restrictions.
We see that the triplet amplitudes can be quite long
ranged in S: this is evident, with our phase convention,  
for the imaginary parts of 
$f_0$ and $f_1$ at $\alpha=40^\circ$.
Thus the triplet correlations for this particular 
magnetization orientation 
can penetrate 
all the way to the other end of the S side, even though
the S layer is three 
coherence lengths thick.
In addition, one can see that antiparallel magnetizations in the F layers lead
to both the real parts and the 
imaginary parts of $f_1$ being short ranged.
Non-collinear relative orientations of the exchange fields
in the inner and outer
F layers may induce both long range $f_0$ and $f_1$ components simultaneously. 
However, the triplet pairing correlations for $\alpha=80^\circ$ are not as long ranged
as those for
$\alpha=40^\circ$. This can be indirectly attributed to much weaker singlet amplitudes inside
S  in the former
case: the overall superconductivity scale is still set by
the singlet, intrinsic correlations. 
Considering now the real parts of $f_0$ and $f_1$, and
other than parallel and antiparallel  magnetizations,
the penetration of $f_0$ correlations within the ferromagnet regions
is weakly dependent
on the angle $\alpha$, while $f_1$ is more sensitive to
$\alpha$.
Within the superconductor we see similar trends as  for the imaginary parts 
except that the 
real components of $f_1$ and $f_0$ extend over a
shorter distance within S at the same $\tau$ value.

Motivated by the long range triplet amplitudes found above for 
an ${\rm F_1 F_2 S}$ structure with relatively thin 
F layers and a thick  S layer,
we  discuss next, in Fig.~\ref{fig4}, the case of a thicker outer ferromagnet
layer with
$D_{F1}=60$, with $D_S=150$. The values of
$\tau$ and $I$ are the same as in Fig.~\ref{fig3}.
Fig.~\ref{fig4} shows that the triplet amplitudes are more prominent in the F 
than in the S regions.  There is also an underlying periodicity 
that is superimposed with apparent 
interference effects, with a shorter period 
than that found in the singlet pair amplitudes (see bottom panel, Fig.~\ref{fig2}).
Also, the imaginary component of $f_0$  
penetrates the superconductor less than the imaginary $f_1$ component. 
For the real $f_1$ component, the exchange
field of the inner layer produces a valley near the interface in
the F regions. This feature is most prominent 
when the 
exchange fields are anti-parallel, in which case the equal-spin triplet
correlations are maximized. Aside from this, 
the triplet amplitudes in S are smaller than
in the case above with thicker S and thinner $F_1$, although their
range is not dissimilar. 
This is mainly because the triplet
penetration  into S is appreciably affected by finite size effects: 
When one of the F layers is relatively thick, it is only after
a longer time delay $\tau$ that the triplet correlations 
evolve. 
From Fig.~\ref{fig4}, one can also
see that the triplet $f_0$ correlations in S are nearly real (i.e. in
phase with the singlet) and 
essentially
independent of the angle 
$\alpha$.

\begin{figure*} 
\includegraphics[width=1\textwidth] {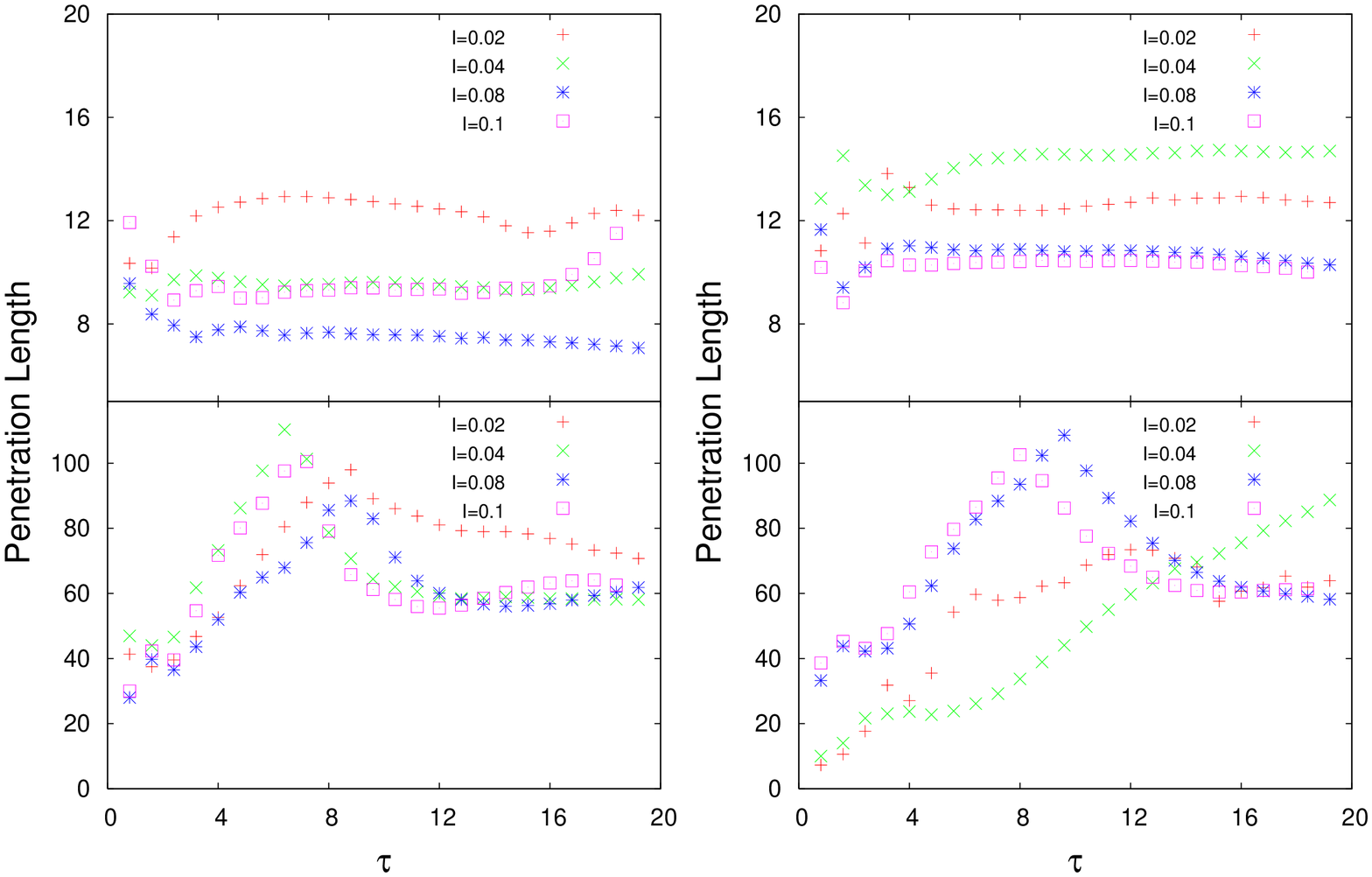}
\caption{(Color online) Study of the triplet penetration
lengths, see Eq.~(\ref{pen}), vs  $\tau$. In this figure, 
$D_F=10$, $D_S=150$, and $\alpha=40^\circ$.
Left panels: lengths as extracted from $f_0(Y)$ 
for several values of $I$ in the $F$ regions (top panel) and in
the $S$ region (bottom panel).
The right panels contain the same information, arranged in
the same way, but with the penetration length extracted from
$f_1(Y)$. The lengths eventually saturate. 
}
\label{fig7}
\end{figure*}

The triplet  penetration is a  function of the characteristic time
$\tau$ scales. We therefore study the dependence of the triplet amplitudes
on $\tau$  in
Fig.~\ref{fig5}, which shows results corresponding to 
$D_F=10$, $D_S=150$, $\alpha=40^\circ$,  $I=0.1$, and at
four 
different values of $\tau$. 
Again, the triplet amplitudes, 
particularly their 
imaginary parts, are long range. 
The plots clearly show that at  short times, 
$\tau=0.8$, the triplet correlations generated at the interface 
reside mainly in the F region. 
At larger values of $\tau$, the triplet amplitudes penetrate
more deeply into the S side, and eventually 
saturate. For the range of times shown, the 
magnitude of the real parts of 
$f_0$ and $f_1$, decays in 
the S region near the interface due to the phase decoherence associated 
with conventional proximity effects. 
For the largest value of $\tau =7.2$  in the figure, the imaginary parts of 
$f_0$ and $f_1$ do not display  monotonic decrease 
on the S side of the interface
but saturate.
This is because for these values of $\tau$ the triplet amplitudes
already pervade the entire S. 
This indicates that 
both triplet components 
infiltrate the superconductor more efficiently
and  at smaller values of $\tau$  
when they are nearly out of phase with the singlet amplitude.

We also investigated the dependence of the triplet amplitudes 
on the magnitude of exchange field at a set time, $\tau=4$. 
Fig.~\ref{fig6} illustrates the real and imaginary parts of
the complex $f_0$ (left panels) and $f_1$ (right panels). 
The geometric parameters are $D_F=10$, and $D_S=150$, 
and we consider four different $I$ values at fixed relative orientation, $\alpha=40^\circ$. 
In our discussion below, we  divide these four different values into two groups, 
the first including the
two smaller values, $I=0.02$ and $I=0.04$,  
and  the second the two somewhat 
larger ones, $I=0.08$ and $I=0.1$.  
In each group, the triplet amplitudes are similar in  shape but different
in magnitude. For the first group, there are no nodes at the 
${\rm F_2S}$ interface for the 
$f_0$ components, while the $f_1$ components 
cross zero near it. 
For the second group,  the opposite occurs: the $f_0$ components cross zero
while the $f_1$ components do not. 
Also, the ratio of ${\rm Re} [f_0]$ 
at $I=0.04$ to  ${\rm Re}[f_0]$ at $I=0.02$ is comparable 
to the ratio for the corresponding singlet amplitudes. 
This can be inferred, see Fig.~\ref{fig1}, 
from  the transition temperatures for $I=0.02$, which are
higher than $I=0.04$. Furthermore, the transition temperatures for the
first group are monotonically increasing with $\alpha$, while for the second 
group, they are non-monotonic functions with a minimum  around $\alpha=80^\circ$.
Therefore, the $f_0$ triplet amplitudes are indeed correlated with singlet amplitudes
and the transition temperatures also reflect their behaviors indirectly.

There is an interesting relationship involving the interplay 
between singlet and equal-spin triplet amplitudes: 
When $T_c$ is a non-monotonic function
of $\alpha$, the singlet amplitudes (which are directly correlated with $T_c$)
at the angle where $T_c(\alpha)$ has a minimum 
are partly transformed into equal spin triplet amplitudes.
By looking at the central panel of Fig.~\ref{fig1}, one sees 
that the transition temperatures 
for  $I=0.08$ and $I=0.1$ nearly overlap, while
the $I=0.02$ case has a much higher transition temperature around $\alpha=80^\circ$. 
The singlet pair amplitudes (at zero temperature) follow the same trend as well: at $I=0.02$,
$F(Y)$ is much larger than the other pair amplitudes at different $I$ (see the 
middle panel of Fig.~\ref{fig2}). The $f_1$ component
for these cases however, shows the opposite trend 
(see, e.g., the right panels of Fig.~\ref{fig6}):
for $I=0.08$ and $I=0.1$, the equal spin correlations extend throughout the S region,
but then abruptly plummet for $I=0.02$.
This inverse relationship between ordinary singlet correlations
and $f_1$ is suggestive of  singlet-triplet conversion
for these particular  magnetizations in each ferromagnet layer.

Having seen that the triplet amplitudes  generated by the inhomogeneous
magnetization can extend  throughout the sample
in a way that depends on $\tau$, we proceed now to characterize
their extension by determining a characteristic triplet proximity length. 
We calculate the characteristic lengths,
$l_i$, from  our data for the  triplet amplitudes, 
by 
using the same definition as in previous\cite{hv2p} work:
\begin{align}
\label{pen}
l_i=\dfrac{\int dy|f_i(y,\tau)|}{\max|f_i(y,\tau)|}, \quad i=0,1,
\end{align}
where the integration is either over the superconducting or 
the magnetic region. The normalization means that these lengths
measure the range, not  the magnitude, of the induced correlations. 
In Fig.~\ref{fig7} we show results for the four
lengths thus obtained, for a sample with $D_F=10$, $D_S=150$ and $\alpha=40^\circ$,
at several values of $I$. 
The left panels show these lengths for the $f_0$ component, 
and the right panels show the results for
the corresponding $f_1$ component. The triplet penetration lengths in the $F$ region
are completely saturated, even at smaller values of $\tau$, 
for both $f_0$ and $f_1$. This saturation follows  only in part from
the relatively thin F layers used for the calculations in this figure: the
same saturation occurs for the geometry of Fig.~\ref{fig4} where
$D_{F1}+D_{F2}=66$, although of course at much larger values of $l_i$. 
The triplet correlations easily pervade the magnetic part of the sample. 
On the other hand, the corresponding penetration lengths for both 
triplet correlations, $f_0$ and $f_1$, in the S region 
are substantially greater and,
because $D_S$ is much larger, do not saturate but
possess a peak around $\tau=8$ in all cases except for $f_1$ at lager $I$
where it is beyond the figure range. The behavior for the sample with
larger F thicknesses is, on the S side, qualitatively similar.




\begin{figure} 
\includegraphics[width=0.5\textwidth] {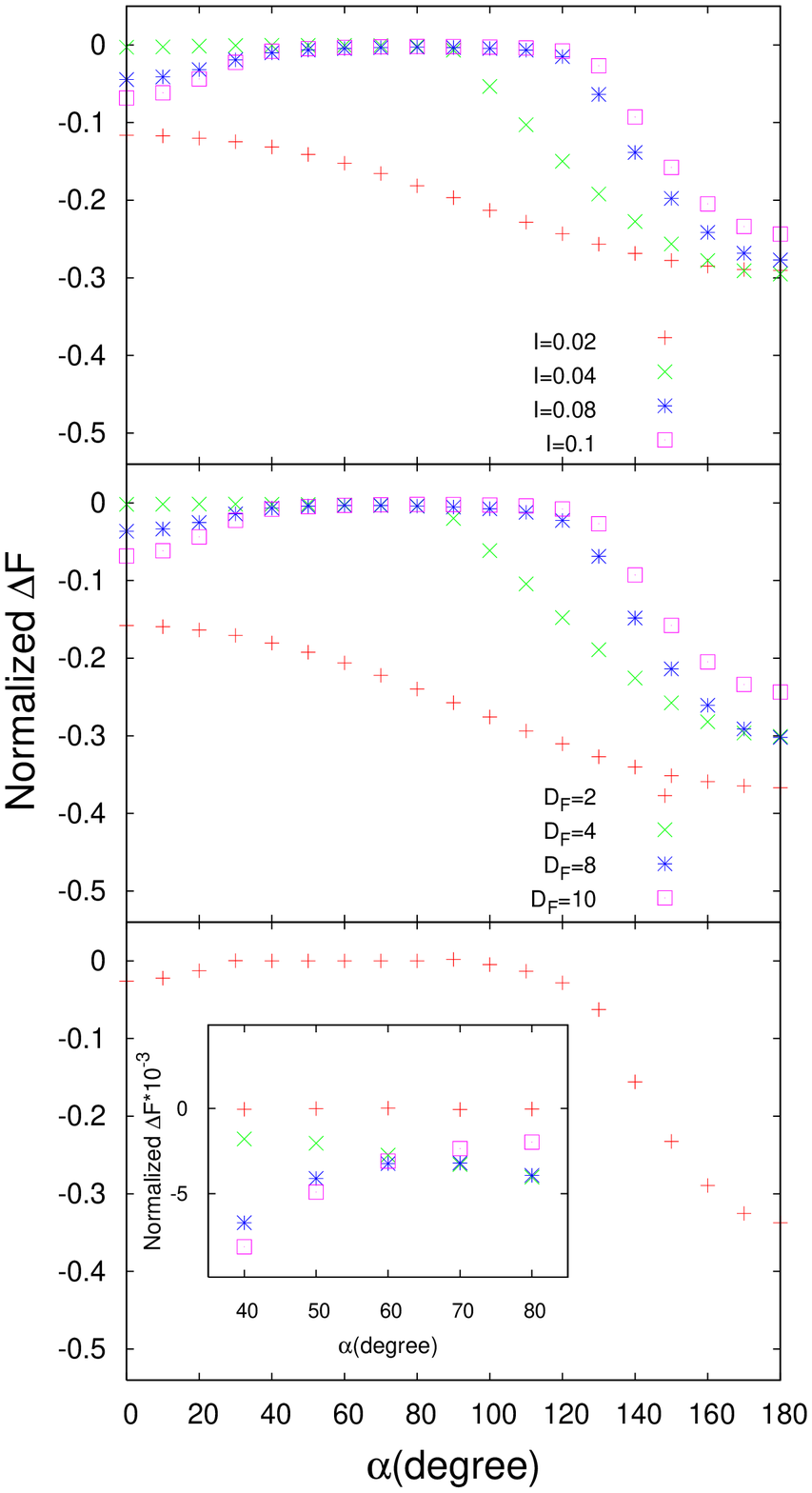} 
\caption{(Color online)  Normalized condensation free energies $\Delta F$
 vs $\alpha$, at $T=0$. 
 The three panels are for the same geometry and parameter
 values as in Fig.~\ref{fig1}, and the symbols have the same meaning.
 Thus the top  panel corresponds to $D_F=10$ and several
 values of $I$, while the middle panel is for $I=0.1$ and
 several values of $D_F$. The bottom panel corresponds to the
 reentrant case shown in the corresponding panel of Fig.~\ref{fig1}.
 The inset shows the difference between truly reentrant cases and those 
 for which the condensation energy is small 
 (see text) in the range of $\alpha=40^{\circ}$ to $\alpha=80^{\circ}$.
}
\label{fig9}
\end{figure}

\subsection{Thermodynamics}

Given the self-consistent solutions, we are able to 
compute also the thermodynamic functions.
In particular, we obtained 
the condensation free energies $\Delta F=F_S-F_N$ by using Eq.~\ref{fe}.
In Fig.~\ref{fig9}, we plot  calculated results for $\Delta F$ 
at zero $T$, equivalent to the condensation energy. 
We normalize $\Delta F$ to $N(0)\Delta_0^2$, where $N(0)$ denotes the density of states at 
the Fermi level and $\Delta_0$ denotes the bulk value of the singlet pair potential
in S: 
thus we would have $\Delta F=-0.5$ for pure bulk S.
The three panels in this figure correspond to those in Fig.~\ref{fig1}.
The geometry is the same and the symbol meanings in each panel correspond
to the same cases, for ease of comparison.
In the top panel, we see that the  $\Delta F$ curves for $I=0.02$ and
$I=0.04$ are monotonically decreasing 
with $\alpha$. This corresponds to the monotonically increasing $T_c$. 
One can conclude that the system
becomes more superconducting when $\alpha$ is changing 
from parallel to anti-parallel:
the superconducting state is getting increasingly more favorable than 
the normal one 
as one increases the tilt from $\alpha=0$ to $\alpha=\pi$. 
The other two curves in this panel, which correspond to $I=0.08$ and $I=0.1$, show a maximum
near $\alpha=80^{\circ}$. Again, this is consistent with the transition temperatures 
shown in Fig.~\ref{fig1}.
Comparing also with the middle panel of Fig.~\ref{fig2}, we 
see that the singlet amplitude for $I=0.02$ is much 
larger than that for the other values of $I$. This is 
consistent with Fig.~\ref{fig9}: $\Delta F$
is more negative at $I=0.02$ and the superconducting state is also more stable. 
The middle panel 
of Fig.~\ref{fig9} shows $\Delta F$ for different ferromagnet thicknesses. 
The curves are very similar
to those in the top panel, just as the top two panels in Fig.~\ref{fig1}
were found to be similar to each other. 
Therefore, both Fig.~\ref{fig1} and Fig.~\ref{fig9}, show
that the superconducting states are  
thermodynamically more stable at $\alpha=180^\circ$ than in the intermediate regions 
($\alpha=40^{\circ}$ to $\alpha=80^{\circ}$).
From the top two panels in
Fig.~\ref{fig9}, we also see that $\Delta F$ at $\alpha=180^\circ$ can be near  
$-0.3$ in this geometry: this is a very
large value, quite comparable to that in pure bulk S.
However, in the region of  the $T_c$ minima near $\alpha =80^\circ$, the
absolute value of the condensation energy can be over 
an order 
of magnitude smaller, although it remains (see below)
negative.  
The bottom panel of Fig.~\ref{fig9} shows $\Delta F$
for the reentrant case previously presented in
Fig.~\ref{fig1}, for which $D_F=6$ and $I=0.15$.
The main plot shows the condensation energy results, which vanish
at intermediate angles.
Because $\Delta F$ in the intermediate non-reentrant  regions shown
in the upper two panels can be
very small, in the vertical scale shown, 
we have added to the lowest panel an inset where
the two situations are contrasted. 
In the inset, the (red) plus signs represent $\Delta F$ for the 
truly reentrant case and the other three symbols have the same meaning 
as in the  middle panel, where no reentrance occurs. 
The inset clearly shows the difference:  
$\Delta F$  vanishes in the intermediate region only for the
reentrant $T_c$ case and remains slightly negative otherwise.
The pair amplitudes for the reentrant  region  
are found self-consistently to be identically zero. 
Thus one can safely say that in the
intermediate region the system must stay in the normal state 
and no self-consistent superconducting solution exists.
Evidence for reentrance with $\alpha$ in $\rm F_1F_2S$ 
is therefore  found from both the microscopic pair amplitude and from $T_c$: 
it is also confirmed thermodynamically. 
That superconductivity in ${\rm F_1F_2S}$ trilayers can be reentrant 
with the angle between $\rm F_1$ and $\rm F_2$ layers, makes 
these systems ideal candidates for spin-valves. 

\begin{figure*}
\includegraphics[width=1\textwidth] {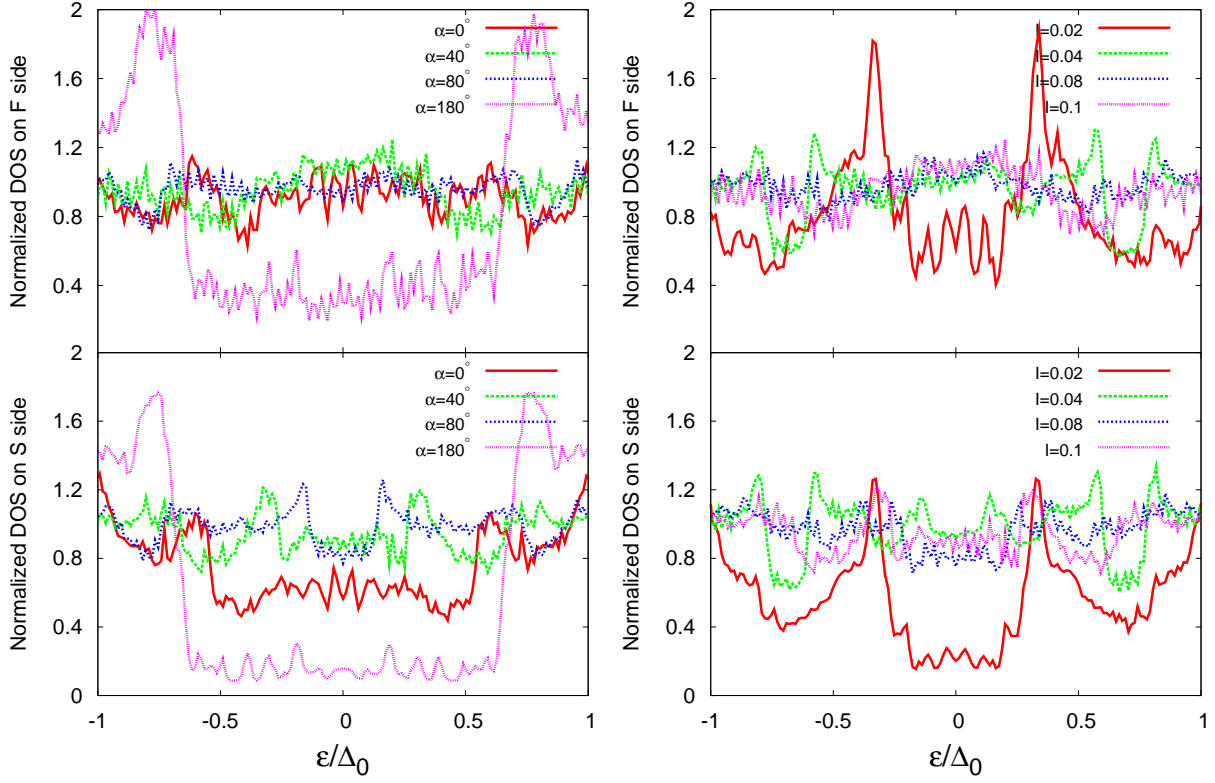}
\caption{(Color online) LDOS integrated over the $F$ layers (top panels) and the
$S$ layer (bottom panels). In all cases 
$D_F=10$, $D_S=150$, and $T=0.05T_c^0$. The left panels show results for $I=0.1$ and the
indicated values of $\alpha$, while in the right panels we have
$\alpha=40^\circ$ and several values of $I$. 
}
\label{fig10}
\end{figure*}

\subsection{DOS} 

Next, we present some results for the local DOS 
(LDOS) in $\rm F_1F_2S$ systems.  All plots are normalized to the
corresponding value in a bulk sample of S material 
in its normal state.
The top panels in Fig.~\ref{fig10} show the normalized LDOS integrated
over the entire magnetic portion of the sample, while in the bottom
panels the LDOS is integrated over the S region. In all four cases
we use $D_F=10$, $D_S=150$. In the left panels we have fixed $I=0.1$
and present results for several angles, while in the right panels we take a
fixed $\alpha=40^\circ$ and show results  for several values of $I$ as indicated.
In the top left panel (F side) we see no energy gap for any value
of $\alpha$, however
a flat valley between two peaks for the case $\alpha=180^{\circ}$ resembles 
a 
characteristic feature of the DOS
in bulk superconductors. However, the plots at the other three angles,
where the transition temperature and condensation energies are much lower,
are very near the value of the DOS in its normal state throughout all energies. 
This is also consistent with the top panel of Fig.~\ref{fig2}, 
where the Cooper pair amplitudes in this case
are larger inside F most significantly at $\alpha=180^{\circ}$.
The singlet amplitudes at $\alpha=0$ are also larger than in the other 
non-collinear configurations, but
the superconducting feature in the LDOS is not as prominent.
This could be due to the contributions from the triplet pairing correlations: 
We know from the spin symmetry arguments discussed
above that there is no $f_1$ component of the induced triplet amplitude 
at $\alpha=0^{\circ}$ and therefore it can not enhance the superconducting feature in 
the DOS.
On the contrary, both singlet and triplet amplitudes can contribute 
when $\alpha=180^{\circ}$.
Thus the LDOS results in the F side reflect 
the signature of induced triplet amplitudes in $\rm F_1F_2S$ 
systems.

The left bottom panel displays the integrated LDOS over the entire S layers for the 
same parameters as the top one. Again, the plot for $\alpha=180^{\circ}$, 
corresponding to the highest $T_c$ and most negative 
condensation energy, possesses a behavior similar
to that 
in pure bulk S material, although the wide dip in the DOS does
not quite reach down to zero. On the other hand, the LDOS at $\alpha=80^{\circ}$, 
the case with the most fragile superconductivity, has a shallow and narrow valley.
The DOS plots on the left side 
are very similar to the normal state
result both at $\alpha=40^{\circ}$ and at $\alpha=80^{\circ}$.
In summary, the depth and the width of the dip are mostly correlated with
the singlet pair amplitudes.
The left panels also support our previous analysis: the slight difference 
between the normal states and superconducting states in the intermediate angle region
is reflected in the DOS. 
The right panels 
reveal
how the magnetic strength parameter, $I$, affects the integrated DOS. 
As we can see from the middle
panel in Fig.~\ref{fig2}, the singlet Cooper pair amplitudes for this case drop
significantly when $I\geq0.04$. The right panels in Fig.~\ref{fig10}
confirm this information, that is
the integrated DOSs in both the F and S sides have a very noticeable dip
in the F side, and a near gap on the S region for $I=0.02$ , 
while for the other vaues of $I$  the eveidence for superconductivity in the
DOS is much less prominent.

\subsection{Local magnetization} 

\begin{figure}
\includegraphics[width=0.5\textwidth] {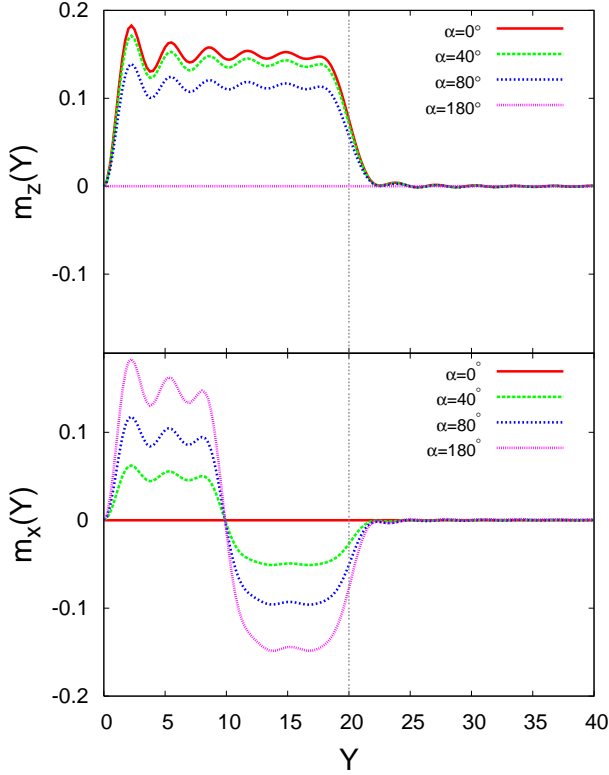}
\caption{(Color online) The $z$ component (top panel) and
the x component (bottom panel)  of the  local magnetization, plotted as a function
of $Y$ for several $\alpha$ values. We use $D_F=10$, $D_S=150$, and $I=0.1$
in this figure.
}
\label{fig11}
\end{figure}

Finally, it is also important to study the reverse proximity effects: 
not only can the superconductivity penetrate into the ferromagnets, but 
conversely the electrons
in S near the interface can be spin polarized by the presence
of the F layers. 
This introduction of magnetic order in S is 
accompanied by a corresponding decrease of the local
local magnetization
in ${\rm F_2}$ near the S interface. 
In Fig.~\ref{fig11},
we show the components of the
local magnetization, as defined in 
Eq.~\ref{mm}.  The parameters used are $D_F=10$, $D_S=150$
and $I=0.1$ and results are shown
for different values of $\alpha$. The  local magnetization results shown are
normalized by
$-\mu_B(N_{\uparrow}+N_{\downarrow})$, where $N_{\uparrow}=k_F^3 (1+I)^{3/2}/6\pi^2$ and
$N_{\downarrow}=k_F^3 (1-I)^{3/2}/6\pi^2$. 
From the figure one sees at once that 
both the sign and average magnitude of the $m_x$ and $m_z$ components inside the F
material are in accordance with the values of the angle $\alpha$ and of 
the exchange field
($I=0.1$). As to the reverse proximity 
effect, we indeed see a nonzero value of the local magnetization in S near the 
the interface. The penetration depth corresponding to this
reverse effect is independent of $\alpha$. 
Unlike the singlet and triplet amplitudes, which may spread 
throughout the entire structure,
the local magnetizations can only penetrate a short distance.
This  is consistent with results from  past work\cite{cite:hv2}. 

\section{Summary and Conclusions} 

In summary, we have investigated 
the proximity effects in ${\rm F_1F_2S}$ trilayers
by self-consistently solving the BdG equations. One of the most prominent features of 
these systems, which  
make them different from  $\rm F_1 S F_2$ structures is the non-monotonicity of $T_c(\alpha)$,
as the angle, $\alpha$, between adjacent magnetizations is varied.
For $\rm F_1SF_2$  systems the critical temperature is always lowest 
for parallel ($\alpha = 0^\circ$) orientations, due chiefly to the 
decreased average exchange field as $\alpha$ increases and 
the two F's increasingly counteract one another.
In contrast, we find 
that $\rm F_1 F_2 S$ configurations can exhibit for
particular combinations of exchange field strengths and layer thicknesses,
critical temperatures that are lowest for relative magnetization orientations 
at an intermediate angle
between the parallel and antiparallel configurations.
In some cases the drop in $T_c$ from the parallel state, as $\alpha$ is varied, is large enough
that superconductivity is completely inhibited over a range of $\alpha$, and then reemerges again
as $\alpha$ increases: the system exhibits
reentrant superconductivity with $\alpha$. 
We also calculated the singlet pair amplitude  and condensation  energies 
at zero temperature, revealing behavior that is entirely  consistent
with these findings. 

We have 
studied  the odd triplet amplitudes that we find 
are generated, and found that both the 
opposite spin pairing (with $m=0$) amplitude,
$f_0$, and the equal-spin pairing amplitude (with $m=\pm 1$), $f_1$, 
can be induced by the inhomogeneous exchange fields in 
the F layers. Also of importance, we have shown 
that the triplet 
pairing correlations 
can be very long ranged and extend throughout both the F and S regions, 
particularly for relatively thick S  and F layers. 
We have characterized this penetration by calculating and analyzing properly
defined characteristic lengths. 
We have also shown 
that the inner $\rm F_2$ layer, when its exchange field is not aligned with
that of
the outer $\rm F_1$ layer, plays an important role in 
generating the triplet amplitudes. 
When both magnets are thin, there is an indirect relationship 
between the singlet pairing amplitudes that govern $T_c$ and
the $f_1$ amplitudes that govern the behavior of equal-spin pairing. 
We have also presented calculations of 
the energy resolved DOS,
spatially averaged over the S or F regions,
demonstrating clear signatures in the energy spectra, which can be identified depending on the
relative magnetization vectors in the $\rm F_1$ and $\rm F_2$ regions.
We have 
determined that the extent of magnetic leakage into the S region
as extracted from a calculation of the components of the  local magnetization,
is
rather short ranged.
Throughout this paper, we have emphasized the potential of these structures as
ideal candidates for spin valves. 

\acknowledgments

We thank C. Grasse and B. Benton for technical help. K.H. 
is supported in part by ONR and by grants
of HPC resources from 
DOD (HPCMP).

\end{document}